\documentclass[12pt]{article}
 \usepackage{graphicx}
 \usepackage[cp1251]{inputenc} 

\begin{document}
 \noindent {\footnotesize\it Astronomy Letters, 2016, Vol. 42, No. 2, pp. 90--99.}
 \newcommand{\dif}{\textrm{d}}

 \noindent
 \begin{tabular}{llllllllllllllllllllllllllllllllllllllllllllll}
 & & & & & & & & & & & & & & & & & & & & & & & & & & & & & & & & & & & & & \\\hline\hline
 \end{tabular}

  \vskip 1.0cm
  \centerline{\bf Kinematic Analysis of Solar-Neighborhood Stars Based on RAVE4 Data}
  \bigskip
  \centerline{V.V. Bobylev and A.T. Bajkova}
  \bigskip
  \centerline{\small\it Pulkovo Astronomical Observatory, St. Petersburg,  Russia}
  \bigskip
  \bigskip
{\bf Abstract}—We consider stars with radial velocities, proper
motions, and distance estimates from the RAVE4 catalogue. Based on
a sample of more than 145000 stars at distances $r<0.5$~kpc, we
have found the following kinematic parameters:
 $(U,V,W)_\odot=(9.12, 20.80, 7.66)\pm(0.10, 0.10, 0.08)$ km s$^{-1}$,
 $\Omega_0=28.71\pm0.63$ km s$^{-1}$ kpc$^{-1}$, and
 $\Omega'_0=-4.28\pm0.11$ km s$^{-1}$ kpc$^{-2}$. This gives the linear
rotation velocity $V_0=230\pm12$ km s$^{-1}$ (for the adopted
$R_0= 8.0\pm0.4$ kpc) and the Oort constants $A=17.12\pm0.45$ km
s$^{-1}$ kpc$^{-1}$ and $B=-11.60\pm0.77$ km s$^{-1}$ kpc$^{-1}$.
The 2D velocity distributions in the $UV,$ $UW,$ and $VW$ planes
have been constructed using a local sample, $r<0.25$~kpc,
consisting of $\sim$47000 stars. A difference of the $UV$ velocity
distribution from the previously known ones constructed from a
smaller amount of data has been revealed. It lies in the fact that
our distribution has an extremely enhanced branch near the Wolf
630 peak. A previously unknown peak at $(U,V)=(-96,-10)$~km
s$^{-1}$ and a separate new feature in the Wolf 630 stream, with
the coordinates of its center being $(U,V)=(30,-40)$ km s$^{-1}$,
have been detected.


\section*{INTRODUCTION}
The distribution of stellar space velocities in the solar
neighborhood has a complex structure. First, characteristic peaks
associated with known open clusters are identified (Skuljan et al.
1999; Famaey et al. 2005; Bobylev and Bajkova 2007; Antoja et al.
2008; Bobylev et al. 2010). Second, a bimodal velocity
distribution is observed. Some authors believe this bimodality to
have a dynamical nature related to the influence of the central
Galactic bar on the motion of stars close to the Suns (Dehnen
1999, 2000; Fux 2001; Chakrabarty 2007; Gardner and Flinn 2010).

Highly accurate space velocities of stars, implying the
availability of their proper motions, radial velocities, and
distances, are required to study these phenomena. Other
characteristics of stars, such as the metallicity indices and age
estimates, are also important. Data from several catalogues have
contributed greatly to the solution of the problem. These include
the proper motions and parallaxes from the Hipparcos catalogue
(ESA 1997; van Leeuwen 2007), the radial velocities from the PCRV
compilation (Gontcharov 2006), and the age estimates and
metallicity indices from the Geneva–Copenhagen survey (Holmberg et
al. 2007, 2009). These data were used to analyze the space
velocities of $\sim$20000 solar-neighborhood stars (Bobylev and
Bajkova 2007; Antoja et al. 2008; Klement et al. 2008; Francis and
Anderson 2009; Bobylev et al. 2010). The next big step is expected
to be made after the implementation (approximately after 2018) of
the GAIA space project, when the parallaxes, radial velocities,
and proper motions will be measured for millions of stars with a
microarcsecond accuracy.

The RAVE (RAdial Velocity Experiment) project plays a significant
role in studying stellar streams. It is a large-scale ground-based
project aimed at determining the radial velocities of faint stars
with an accuracy of $\sim$1--2 km s$^{-1}.$ Observations in the
southern hemisphere started in 2003. Four releases of this
catalogue (DR1--DR4) have been published Since then.

Analysis of the RAVE DR1 data (Steinmetz et al. 2006) revealed a
previously unknown stream in the region of ``fast'' stars
($V\approx-160$ km s$^{-1}$) with an age of $\sim$13 Gyr (Klement
et al. 2008). It was designated as KFR08 (Klement et al. 2008).
Having analyzed the RAVE DR2 (Zwitter et al. 2008) and RAVE DR3
(Siebert et al. 2011) data, for which the distances were estimated
by Burnett et al. (2011), Antoja et al. (2012) detected a
previously unknown stream at $(U,V) = (92, -22)$ km s$^{-1}$.

The errors in the spectrophotometric distances of stars from the
RAVE catalogue are about 30\% or more. Nevertheless, various
authors have shown that a number of important kinematic parameters
can be estimated quite reliably based on RAVE data
(Co\c{c}kuno\u{g}lu et al. 2011; Pasetto et al. 2012; Binney et
al. 2014b).

The goal of this paper is to test the space velocities of stars
from the latest version, RAVE DR4 (Kordopatis et al. 2013). This
catalogue contains the original radial velocities, proper motions,
and photometric distances for 425561 stars. We want to examine how
well the Galactic rotation parameters can be determined from
distant stars and to analyze the 2D velocity distributions in the
$UV,$ $UW,$ and $VW$ planes for nearby stars.

\section*{METHODS}
From observations we know three projections of the stellar
velocity: the radial velocity $V_r$ as well as the two velocity
components $V_l = 4.74r\mu_l \cos b$ and $V_b = 4.74r\mu_b$
directed along the Galactic longitude $l$ and latitude $b$ and
expressed in km s$^{-1}.$ Here, the coefficient 4.74 is the ratio
of the number of kilometers in an astronomical unit to the number
of seconds in a tropical year, and $r$ is the star's heliocentric
distance in kpc. The proper motion components $\mu_l \cos b$ and
$\mu_b$ are expressed in milliarcseconds per year (mas yr$^{-1}$).
The velocities $U,$ $V,$ and $W$ directed along the rectangular
Galactic coordinate axes are calculated via the components $V_r,$
$V_l,$ and $V_b:$
 \begin{equation}
 \begin{array}{lll}
 U=V_r\cos l\cos b-V_l\sin l-V_b\cos l\sin b,\\
 V=V_r\sin l\cos b+V_l\cos l-V_b\sin l\sin b,\\
 W=V_r\sin b                +V_b\cos b,
 \label{UVW}
 \end{array}
 \end{equation}
where $U$ is directed from the Sun to the Galactic center, $V$ is
in the direction of Galactic rotation, and $W$ is directed toward
the north Galactic pole.

To determine the parameters of the Galactic rotation curve, we use
the equations derived from Bottlinger’s formulas, in which the
angular velocity $\Omega$ is expanded in a series to terms of the
second order of smallness in $r/R_0:$
\begin{equation}
 \begin{array}{lll}
 V_r=-U_\odot\cos b\cos l-V_\odot\cos b\sin l\\
 -W_\odot\sin b+R_0(R-R_0)\sin l\cos b\Omega^\prime_0
 +0.5R_0(R-R_0)^2\sin l\cos b\Omega^{\prime\prime}_0,
 \label{EQ-1}
 \end{array}
 \end{equation}
 \begin{equation}
 \begin{array}{lll}
 V_l= U_\odot\sin l-V_\odot\cos l-r\Omega_0\cos b\\
 +(R-R_0)(R_0\cos l-r\cos b)\Omega^\prime_0
 +0.5(R-R_0)^2(R_0\cos l-r\cos b)\Omega^{\prime\prime}_0,
 \label{EQ-2}
 \end{array}
 \end{equation}
 \begin{equation}
 \begin{array}{lll}
 V_b=U_\odot\cos l\sin b + V_\odot\sin l \sin b\\
 -W_\odot\cos b-R_0(R-R_0)\sin l\sin b\Omega^\prime_0
    -0.5R_0(R-R_0)^2\sin l\sin b\Omega^{\prime\prime}_0,
 \label{EQ-3}
 \end{array}
 \end{equation}
Here, $R$ is the distance from the star to the Galactic rotation
axis:
  \begin{equation}
 R^2=r^2\cos^2 b-2R_0 r\cos b\cos l+R^2_0.
 \end{equation}
In this paper, we take the Galactocentric distance of the Sun to
be $R_0=8.0\pm0.4$~kpc, as estimated by Foster and Cooper (2010).

To identify statistically significant signals from the main clumps
in the $UV$ velocity distributions, we use a wavelet transform,
which is known as a powerful tool for filtering spatially
localized signals (Chui 1997; Vityazev 2001).

The wavelet transform of a 2D distribution $f(U,V)$ consists in
its decomposition into analyzing wavelets $\psi(U/a,V/a)$, where
$a$ is the coefficient that allows a wavelet of a certain scale to
be separated from the entire family of wavelets characterized by
the same shape $\psi$. The wavelet transform $w(\xi,\eta)$ is
defined as a correlation function in such a way that at any given
point $(\xi,\eta)$ in the $UV$ plain, we have one real value of
the following integral:
  \begin{equation}
w(\xi,\eta)=\int_{-\infty}^\infty \int_{-\infty}^\infty f(U,V)
\psi\Biggl(\frac{(U-\xi)}{a},\frac{(V-\eta)}{a}\Biggr) dU dV,
 \end{equation}
called the wavelet coefficient at point $(\xi,\eta)$. Obviously,
in the case of finite discrete maps that we are dealing with,
their number is finite and equal to the number of square bins on
the map.

As an analyzing wavelet we use the traditional wavelet called the
Mexican Hat (MHAT). The 2D MHAT wavelet is described by the
expression
\begin{equation}
 \psi(d/a)=\Biggl(2-\frac{d^2}{a^2}\Biggr)e^{-d^2/2a^2},
 \label{w4}
 \end{equation}
where $d^2=U^2+V^2$. The wavelet (7) is obtained by
differentiating the Gaussian function twice. The main property of
the wavelet $\psi$ is that its integral over $U$ and $V$ is zero,
which allows any clumps in the distribution being studied to be
detected. If the distribution being analyzed is uniform, then all
coefficients of the wavelet transform will be zero.

\section*{DATA}
Following Williams et al. (2013), we take the stars satisfying the
following criteria to select candidates without significant random
observational errors:
\begin{equation}
 \begin{array}{ccc}
 |V_r|<600~\hbox {km s$^{-1}$}, \\
 |\mu_\alpha, \mu_\delta|<400~\hbox {mas yr$^{-1}$},\\
 |e_{\mu_\alpha}, e_{\mu_\delta}|<20~\hbox {mas yr$^{-1}$}.
\label{cut-1}
 \end{array}
 \end{equation}
In the RAVE DR4 catalogue containing the proper motions from
various sources, we used those copied from the UCAC4 catalogue
(Zacharias et al. 2013).

As the RAVE program was being performed, several versions of
spectrophotometric distance estimates were published (Breddels et
al. 2010; Zwitter et al. 2010; Burnett and Binney 2010; Burnett et
al. 2011; Binney et al. 2014a). From the catalogue we take the
distances expressed in kpc (there are also the parallaxes and
distance moduli) and determined by Binney et al. (2014a).
According to the estimates by these authors, the distance errors
are 15--20\% for hot dwarfs, they reach 20--30\% for the coolest
dwarfs, and these errors are even larger for giants.

 \begin{figure} {\begin{center}
 \includegraphics[width=120.0mm]{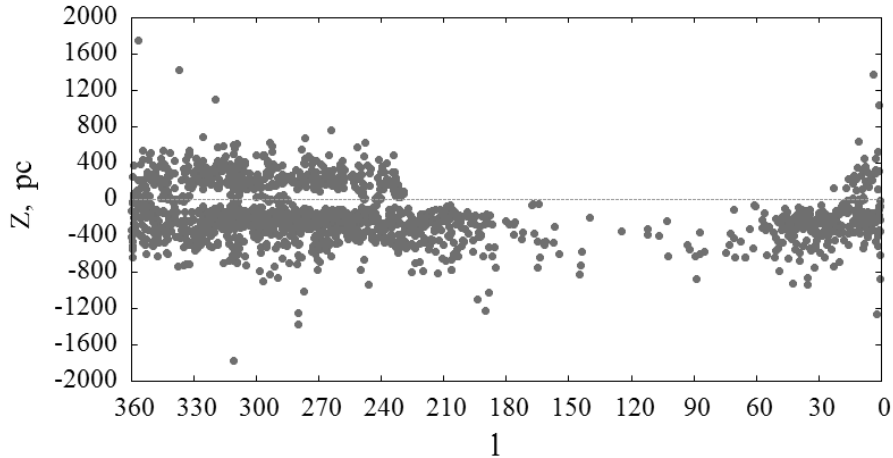}
 \caption{Distribution of a small sample of stars from the RAVE DR4
 catalogue perpendicular to the Galactic plane }
  \label{f1} \end{center} }
 \end{figure}

The technique for estimating the individual ages of stars from the
RAVE DR4 catalogue is described in Burnett and Binney (2010) and
Binney et al. (2014a). This technique consists in comparing the
positions of stars on the Hertzsprung-Russell diagram with
suitable theoretical isochrones. The technique was applied both to
main-sequence stars and to giants. The Bayesian approach is used
for this purpose; it consists in simultaneously seeking for six
unknowns: the metal abundance, the stellar age, the stellar mass,
the heliocentric distance of the star, and the Galactic
coordinates of the star. Such a problem is solved by the maximum
likelihood method. The solution requires (a) observational data
and (b) a priori information, i.e., a model. The Padova isochrones
are ultimately used to determine the ages. As was noted by Burnett
and Binney (2010) and Binney et al. (2014a), the distances and
ages for giants are estimated with the largest errors compared to
those for dwarfs. Kordopatis et al. (2013) point out that the age
and distance estimates obtained for stars from the RAVE4 catalogue
are of interest only from a statistical point of view.

A peculiarity of the RAVE catalogue is a highly nonuniform spatial
distribution of stars, because only the southern hemisphere of the
celestial sphere is observed. The distribution of a small
 $(8.5<\lg t<8.75)$ sample of stars from the RAVE DR4 catalogue
perpendicular to the Galactic plane is shown in Fig.~1. As can be
seen from the figure, a significant space in the range of
longitudes $30^\circ-240^\circ$ remains unfilled in the northern
Galactic hemisphere.

As is well known (Antoja et al. 2012; Williams et al. 2013), the
spatial distribution of stars from the RAVE catalogue resembles a
cone with its axis along the Galactic $Z$ axis. For example, only
a few stars very close to the Sun are located below approximately
150 pc. This effect is also seen in Fig.~1 as a poorly filled band
extended along the equator. Thus, the constraints on $Z$ play an
important role in selecting the necessary stars from the RAVE
catalogue.

 \begin{table}[t]                                     
 \caption[]{\small
 Galactic rotation parameters found from five samples from the distance
 range $r$:~0.5--3~kpc
 }
  \begin{center}  \label{t:01}
  \small
  \begin{tabular}{|l|r|r|r|r|r|}\hline
   Parameters                   &  $J-K<0.5$ & $J-K\geq0.5$  &  $\lg t<9.4$ & $\lg t$: 9.4--9.8 & $9.8<\lg t$ \\\hline
                                &\multicolumn{2}{c|}{$|Z|<0.3$~kpc}&\multicolumn{3}{|c|}{$|Z|<0.6$~kpc} \\\hline
   $U_\odot,$  km s$^{-1}$      & $11.82\pm0.48$ & $13.48\pm0.42$ & $11.58\pm0.37$ & $13.01\pm0.20$ & $12.16\pm0.43$ \\
   $V_\odot,$  km s$^{-1}$      & $18.47\pm0.36$ & $22.02\pm0.24$ & $15.60\pm0.28$ & $22.94\pm0.16$ & $31.35\pm0.36$ \\
   $W_\odot,$  km s$^{-1}$      & $ 8.34\pm0.29$ & $ 6.09\pm0.19$ & $ 7.19\pm0.23$ & $ 6.87\pm0.12$ & $ 7.76\pm0.36$ \\
 $\Omega_0,$   km s$^{-1}$ kpc$^{-1}$  & $23.52\pm0.69$ & $20.40\pm0.37$ & $23.25\pm0.44$ & $22.07\pm0.20$ & $22.99\pm0.39$ \\
 $\Omega^{'}_0,$ km s$^{-1}$ kpc$^{-2}$ & $-3.14\pm0.12$ & $-3.15\pm0.06$ & $-3.05\pm0.08$ & $-3.05\pm0.04$ & $-2.48\pm0.08$ \\
$\Omega^{''}_0,$ km s$^{-1}$ kpc$^{-3}$ & $ 0.79\pm0.29$ & $ 1.06\pm0.12$ & $ 0.33\pm0.16$ & $ 0.51\pm0.07$ & $ 0.33\pm0.11$ \\
   $\sigma_0,$   km s$^{-1}$    &          36.0  &          36.8  &          33.6  &          38.4  &          48.0  \\
     $N_\star$                  &         15933  &         38896  &         21202  &         96107  &         24842  \\
            $V_0,$ km s$^{-1}$  &      $188\pm11$ &     $163\pm9$  &     $186\pm10$ &      $177\pm9$ &      $184\pm9$ \\
    $A,$ km s$^{-1}$ kpc$^{-1}$ & $ 12.56\pm0.48$ & $12.60\pm0.25$ & $ 12.18\pm0.34$ & $12.19\pm0.16$ & $-9.91\pm0.31$\\
    $B,$ km s$^{-1}$ kpc$^{-1}$ & $-10.96\pm0.84$ & $-7.79\pm0.45$ & $-11.06\pm0.55$ & $-9.88\pm0.07$ & $-13.08\pm0.49$\\
 \hline
 \end{tabular}\end{center}  \end{table}

 \begin{table}[t]                                     
 \caption[]{\small
 Galactic rotation parameters found from samples of stars with various ages taken from
 the distance range $r<3$~kpc and $|Z|<0.6$~kpc}
  \begin{center}  \label{t:02}
  \small
  \begin{tabular}{|l|r|r|r|r|r|}\hline
   Parameters                & $\lg t$: 8.5--9.0 &  $\lg t$: 9.0--9.4 & $\lg t$: 9.4--9.8 & $9.8<\lg t$ \\\hline

   $U_\odot,$    km s$^{-1}$    & $10.84\pm0.36$ &  $10.66\pm0.22$ & $10.46\pm0.11$ & $ 9.77\pm0.18$ \\
   $V_\odot,$    km s$^{-1}$    & $13.05\pm0.33$ &  $14.71\pm0.19$ & $21.48\pm0.09$ & $25.73\pm0.17$ \\
   $W_\odot,$    km s$^{-1}$    & $ 6.90\pm0.28$ &  $ 7.17\pm0.16$ & $ 7.19\pm0.08$ & $ 8.14\pm0.16$ \\
 $\Omega_0,$    km s$^{-1}$ kpc$^{-1}$  & $25.21\pm0.59$ &  $23.98\pm0.34$ & $24.36\pm0.13$ & $24.53\pm0.34$ \\
 $\Omega^{'}_0,$ km s$^{-1}$ kpc$^{-2}$ & $-3.76\pm0.12$ &  $-3.17\pm0.07$ & $-3.33\pm0.03$ & $-2.66\pm0.08$ \\
$\Omega^{''}_0,$ km s$^{-1}$ kpc$^{-3}$ & $ 0.00\pm0.25$ &  $-0.08\pm0.14$ & $ 0.11\pm0.05$ & $-0.65\pm0.11$ \\
   $\sigma_0,$   km s$^{-1}$    &          26.8  &           30.7  &          35.2  &          40.1  \\
     $N_\star$                  &          8975  &          35385  &        183631  &         59741  \\
           $V_0,$ km s$^{-1}$   &      $202\pm12$ &      $192\pm10$ &      $195\pm10$ &      $196\pm10$ \\
    $A,$ km s$^{-1}$ kpc$^{-1}$ & $ 15.05\pm0.49$ & $ 12.66\pm0.29$ & $ 13.31\pm0.12$ & $ 10.65\pm0.30$ \\
    $B,$ km s$^{-1}$ kpc$^{-1}$ & $-10.17\pm0.77$ & $-11.32\pm0.44$ & $-11.05\pm0.18$ & $-13.87\pm0.45$ \\
 \hline
 \end{tabular}\end{center} \end{table}

\section*{RESULTS}
\subsection*{Galactic Rotation}
To determine the Galactic rotation parameters, we produced several
samples within 3~kpc of the Sun for solving the system of
equations (2)--(4). The upper boundary of the interval $r=3$~kpc
was chosen here in such a way that the errors of the stellar
proper motions (they increase with distance when converted to km
s$^{-1}$) did not dominate in the sample. The results are
reflected in Tables 1 and 2.

Table 1 presents the results obtained from stars with heliocentric
distances $r$ from 0.5 to 3 kpc. Here, we did not use any stars at
$r<0.5$~kpc to reduce the influence of old low-mass dwarfs, which
are represented in the catalogue mostly at close distances
(Fig.~12 in Binney et al., 2014a). Columns 2 and 3 of the table
present the results obtained from two samples that were selected
under the condition $|Z|<0.3$~kpc and that differ in $(J-K)$ color
index. Note that the $(J-K)$ color index is a temperature index,
while $(J-K)=0.5$ corresponding to a temperature
$T_{eff}\approx5000$~K. Columns 4--6 of the table present the
results obtained from three samples that were selected under the
condition $|Z| < 0.6$~kpc and that differ in age.

As can be seen from the table, there are only $\sim$55000 stars in
the zone $r=0.5-3$~kpc, $|Z|<0.3$~kpc. The parameters found from
these two samples have virtually no differences between
themselves. There are already more stars in the zone
$r=0.5-3$~kpc, $|Z|<0.6$~kpc; therefore, we used three samples of
different ages.

We can see that the components of the Sun's peculiar velocity
$U_\odot, V_\odot, W_\odot$ are determined well from all samples.
$U_\odot$ and $W_\odot$ remain almost constant. $V_\odot$
increases with age of the sample stars, reflecting the well-known
asymmetric drift effect. The second derivative of the angular
velocity of Galactic rotation $\Omega''_0$ is determined well.

The lower rows of Table 1 give the linear velocity of the solar
neighborhood around the Galactic center $V_0=R_0\Omega_0$, along
with the Oort constants $A=0.5R_0\Omega^{'}_0$ and
$B=\Omega_0+0.5R_0\Omega^{'}_0$. The table gives the error per
unit weight $\sigma_0$. This quantity characterizes the dispersion
of the residuals when the system of equations (2)--(4) is solved
by the least-squares method. Its value is usually close to the
dispersion of the residual velocities for the sample stars
averaged over all directions (the ``cosmic'' velocity dispersion)
and correlates well with the age of the sample stars.

Table 2 gives the Galactic rotation parameters found by solving
the system of equations (2)--(4) for four samples of stars with
various ages at $r<3$~kpc under the condition $|Z|<0.6$~kpc. It
can be seen from the number of stars (287732) that the bulk of the
RAVE DR4 catalogue is covered.

When working on the sample of youngest stars, we noticed that the
error per unit weight $\sigma_0$ increased significantly at an age
$\lg t<8.5,$ reaching more than 35 km s$^{-1}.$ Therefore, we do
not use the age interval $\lg t<8.5;$ however, there are very few
such stars.

In general, we see that including nearby ($r<0.5$~kpc) stars led
to an increase in $\Omega_0$ (and, accordingly, in $V_0$), but the
second derivative of the angular velocity of Galactic rotation
$\Omega''_0$ creased to be determined. Since Eqs. (2)--(4) contain
the distance squared at this unknown, $\Omega''_0$ can be
determined only at large sample radii, $r>2$~kpc. Therefore, using
stars at $r<0.5$~kpc, of course, gives nothing for the
determination of $\Omega''_0$. However, in the solution for the
age interval $\lg t=9.4-9.8,$ all of the parameters being
determined have the smallest errors (compared to other age
intervals), and even the sign of $\Omega''_0$ is positive.

To analyze the 2D velocity distributions, we must correct the
velocities $U$ and $V$ for the Galactic differential rotation.
However, since we take stars close to the Sun, no farther than
500~pc, for this purpose, it will suffice to use only two
parameters, for example, $\Omega_0$ and $\Omega'_0$ or the two
Oort constants $A$ and $B,$ to take into account the Galactic
rotation.

To this end, we took 145807 stars at $r<0.5$~kpc and found a
solution of the system of equations (2)--(4) with five unknowns
(without $\Omega''_0$):
 \begin{equation}
 \label{Best_solution}
 \begin{array}{lll}
 (U_\odot,V_\odot,W_\odot)=(9.12,20.80,7.66)\pm(0.10,0.10,0.08)~\hbox {km s$^{-1}$},\\
      \Omega_0 =~28.71\pm0.63~\hbox {km s$^{-1}$ kpc$^{-1}$},\\
     \Omega'_0 =-4.28\pm0.11~\hbox{km s$^{-1}$ kpc$^{-2}$},
 \end{array}
 \end{equation}
This gives the linear rotation velocity of the Sun around the
Galactic center $V_0 = 230\pm12$ km s$^{-1}$ for the adopted
$R_0=8.0\pm0.4$~kpc as well as $A=17.12\pm0.45$ km s$^{-1}$
kpc$^{-1}$ and $B=-11.60\pm0.77$ km s$^{-1}$ kpc$^{-1}$. Based on
this solution, below we correct the sample of local stars for the
Galactic rotation.

 \begin{figure} {\begin{center}
 \includegraphics[width=100.0mm]{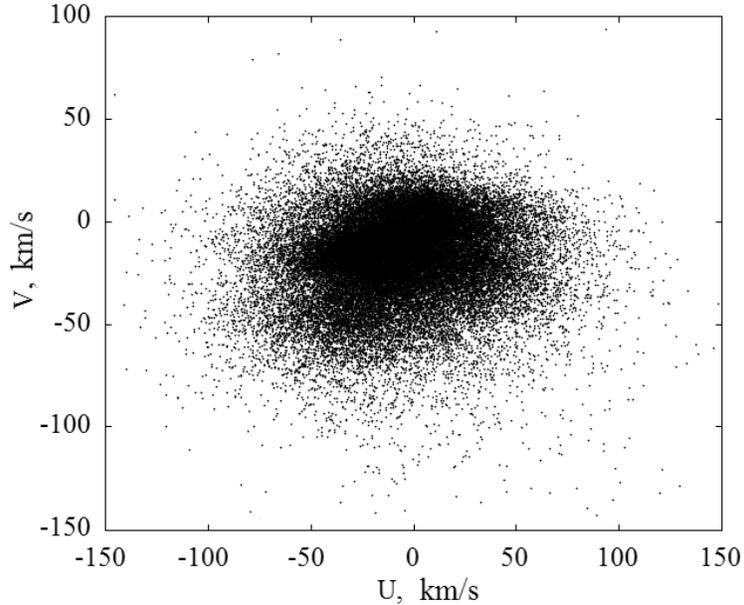}
 \caption{$UV$ velocity distribution for a sample of 47304 stars within 250 pc of the Sun.
 The velocities are given relative to the Sun.}
  \label{f2} \end{center} }
 \end{figure}
 \begin{figure} {\begin{center}
 \includegraphics[width=80.0mm]{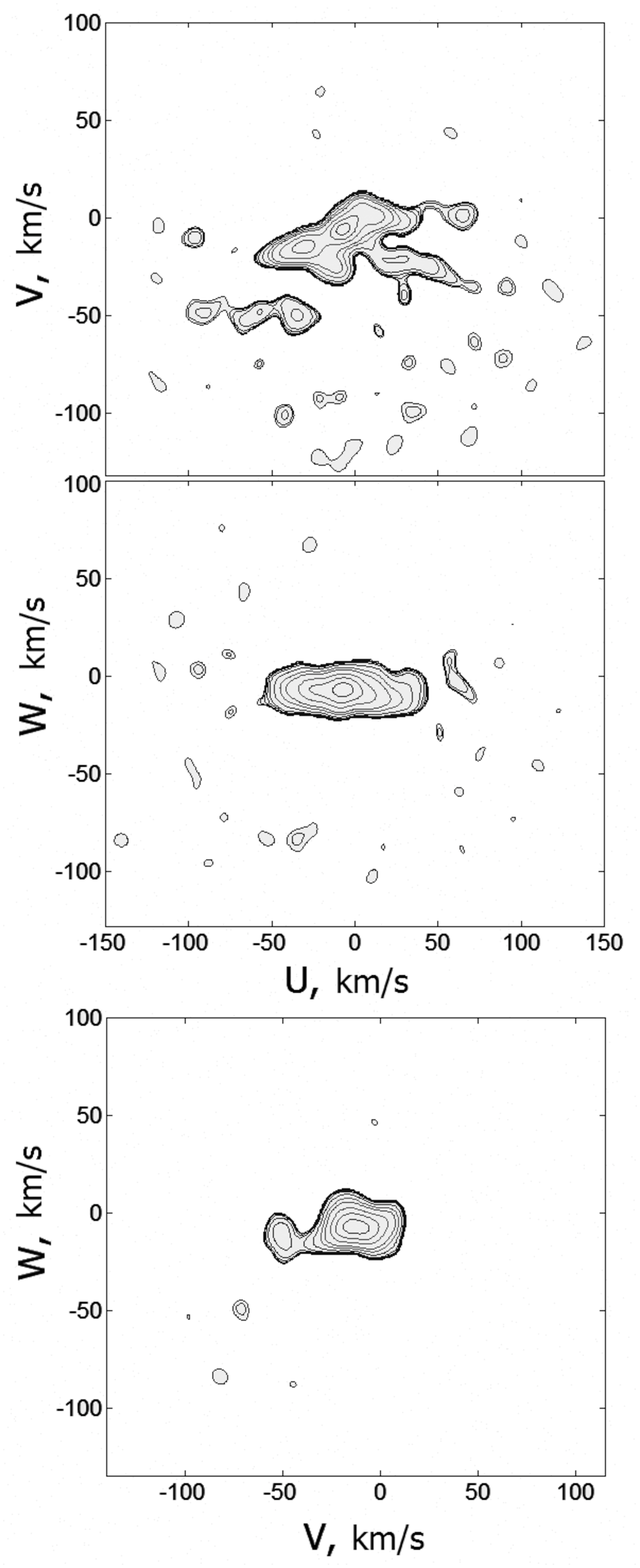}
 \caption{Smoothed wavelet maps of $UV,$ $UW,$ and $VW$ velocities for a sample of
 47304 stars within 250 pc of the Sun. The velocities are given relative to the Sun.}
  \label{f3} \end{center} }
 \end{figure}

\subsection*{Velocity Distributions}
To analyze the 2D velocity distributions, we took stars close to
the Sun. These stars are located within 250 pc of the Sun. $U, V,
W$ are the residual velocities, i.e., they were corrected for the
Galactic rotation with the parameters from solution (9). In
addition, the random errors for each of the velocities $U, V, W$
do not exceed 30 km s$^{-1}$. The $UV$ velocity distribution for
the stars of this sample containing 47304 stars is shown in Fig.
2.

Figure 3 presents the wavelet maps of $UV,$ $UW,$ and $VW$
velocities for our sample of 47304 stars. The following system of
levels on the graphs was chosen: 0.0025, 0.005, 0.01, 0.02, 0.04,
0.1, 0.2, 0.3, 0.4, 0.5, 0.7, and 0.9.

Table 3 gives the coordinates of the main peaks in the $UV$
velocity plane and an approximate number of stars ($N_\star$) in
each of the peaks. The identification with the peaks from Antoja
et al. (2012) is reflected in the last column of the table. Note
that the random errors of the peak centers are approximately equal
in our paper and in Antoja et al. (2012). As follows from Table 2
in Antoja et al. (2012), these errors increase dramatically as one
recedes from the distribution center. For example, the errors in
each of the coordinates are 0.1--0.2 km s$^{-1}$ at the point with
coordinates $(U,V)=(0,0)$ km s$^{-1}$, while toward the map
boundaries, at velocities $U\approx120$ km s$^{-1}$ or
$V\approx120$ km s$^{-1}$, the errors reach $\sim$10 km s$^{-1}$.
Therefore, the classical peaks at the map center, which have their
own names, for example, Pleiades, Hyades, or Sirius, are easily
identified. Closer to the map edges, the discrepancy in peak
coordinates becomes increasingly significant, making the
identification difficult.

For example, the coordinates of stream no. 12 are
$(U,V)=(92,-22)\pm(3.5,1.4)$ km s$^{-1}$ (Antoja et al. 2012). We
identified it with the feature at $(U,V)=(91,-35)\pm(4.2,1.5)$ km
s$^{-1}$ on our map, where this isolated feature is clearly seen
and has two contours. Thus, there is no doubt whatsoever that the
identification is correct.

The coordinates of stream no. 15 are $(U,V)=(60,-72)\pm(3.3,4.7)$
km s$^{-1}$ (Antoja et al. 2012). We identify it with the feature
at $(U,V)=(72,-64)\pm(3.2,3.0)$ km s$^{-1}$ on our map. Given the
errors, the coincidence is seen to be close.

One would think that there should be no difference between the
maps from Antoja et al. (2012) and our maps—in both cases, the
data were taken from the RAVE catalogue after all. However,
different systems of distances were used in these two cases,
leading to differences in the observed velocities $U, V, W.$

Note that all of the isolated peaks with at least two contours on
the $UV$ velocity map (Fig.~3) are reflected in one form or
another in Table~3. If, however, the peak on the $UV$ velocity map
(Fig.~3) has only one contour (quite a low significance), then we
did not include it in the table. The weak peak no.~24 that we
identified with peak no.~13 from Antoja et al. (2012) constitutes
an exception. As follows from Fig.~3 and Table~3, a chain of four
peaks is clearly seen in the region of the Hercules stream: nos.
5, 8, 10, and 24. The situation is similar in the region of the
Wolf~630 stream traced by five peaks: nos. 6, 7, 12, 14, and 17.

 \begin{table}[t]                                     
 \caption[]{\small Coordinates of the main peaks in the $UV$ velocity plane}
  \begin{center}  \label{t:03}
  \small
  \begin{tabular}{|r|r|r|r|l|l|}\hline
  No & $U$ & $V$  & $N_\star$ & Name  \\\hline
     &\multicolumn{2}{c|}{km s$^{-1}$} & &\\\hline

   1 & $- 7$ &  $- 6$ & 1800 & Coma Berenices \\
   2 & $-30$ &  $-15$ & 1500 & Hyades   \\
   3 & $ 10$ &  $  1$ & 1290 & Sirius   \\
   4 & $-13$ &  $-24$ & 1480 & Pleiades \\
   5 & $-35$ &  $-50$ &  390 & Hercules II \\
   6 & $ 29$ &  $-21$ &  624 & Wolf 630  \\
   7 & $ 43$ &  $-24$ &  290 & Dehnen98 (No14)\\
   8 & $-57$ &  $-48$ &  248 & Hercules I \\
   9 & $ 65$ &  $  1$ &  272 & $\gamma$ Leo \\
  10 & $-90$ &  $-49$ &  108 & $\varepsilon$ Ind \\
  11 & $-42$ & $-101$ &  115 & $\eta$ Cep \\
  12 & $ 20$ &  $-22$ &  450 & Dehnen98 (No6) \\
  13 & $ 14$ &  $-58$ &   75 & HR~1614 \\
  14 & $ 30$ &  $-40$ &  395 &   \\
  15 & $-96$ &  $-10$ &  190 &   \\
  16 & $ 91$ &  $-35$ &   40 & Antoja12 (No12) \\
  17 & $ 71$ &  $-35$ &  400 &   \\
  18 & $ 72$ &  $-64$ &   25 & Antoja12 (No15) \\
  19 & $ 35$ &  $-99$ &   99 &   \\
  20 & $ 32$ &  $-74$ &   85 &   \\
  21 & $-10$ &  $-82$ &   83 &   \\
  22 & $-61$ &  $-72$ &   39 &   \\
  23 & $ 84$ &  $-74$ &   49 &   \\
  24 & $-120$ & $-32$ &   42 & Antoja12 (No13) \\
 \hline
 \end{tabular}\end{center} \end{table}

\section*{DISCUSSION}
First of all, it should be noted that the components of the Sun's
group velocity we found, $(U,V,W)_\odot$, agree well with the
results of other authors.

Having analyzed the RAVE DR3 stars that belong to the Galactic
thin disk with a high probability, Co\c{c}kuno\u{g}lu et al.
(2011) found $(U,V,W)_\odot=(8.50, 13.38, 6.49) \pm (0.29, 0.43,
0.26)$ km s$^{-1}$. They should be compared with the results in
the second ($\lg t=8.5-9.0$) or third ($\lg t=9.0-9.4$) columns of
Table 2. There is also good agreement with the results of a
detailed analysis of the velocities $(U,V,W)_\odot$ based on RAVE
data (Pasetto et al. 2012; Karaali et al. 2014). Of interest is
Fig. 7 from Karaali et al. (2014), where the velocity $V_\odot$ is
plotted against the $Z$ coordinate, with the minimum value of this
velocity being $\sim$10 km s$^{-1}$ at $Z=0$~kpc. Pasetto et al.
(2012) concluded that two velocities, $U_\odot$ and $W_\odot$,
remained virtually unchanged at various constraints imposed on the
sample stars. These authors found $(U,W)_\odot=(10.9,7.2) \pm
(1.0,1.3)$ km s$^{-1}.$ The velocity $V_\odot$ is known to reflect
the lag of the centroids behind the Sun (asymmetric drift);
therefore, this velocity increases with increasing age of the
sample stars. According to present-day estimates (Sch\"onrich et
al. 2010; Bobylev and Bajkova 2014), the minimum value of this
velocity is $V_\odot\approx12$ km s$^{-1}.$

At present, there is a sample of about 100 masers whose
trigonometric parallaxes were measured by VLBI with a very high
accuracy, with a mean error of $\pm20$ microarcseconds and, some
of them, with a record errors of $\pm5$ microarcseconds. Having
analyzed these masers, Reid et al. (2014) found the solar velocity
to be $V_0=240\pm8$ km s$^{-1}$ ($R_0=8.34\pm0.16$ kpc). Based on
a smaller number of masers, Honma et al. (2012) obtained an
estimate of $V_0=238\pm14$ km s$^{-1}$ ($R_0=8.05\pm0.45$~kpc).
This is the velocity $V_0$ for the youngest fraction of the
Galactic disk, because the masers used are associated with young
massive protostars. The parameters of solution (9) agree with
these estimates.

The clumps in the $UV$ velocity plane have been repeatedly
analyzed by various authors using the parallaxes and proper
motions from the Hipparcos catalogue (ESA 1997) and, occasionally,
with the inclusion of stellar radial velocities. Note, for
example, the papers by Dehnen (1998), Skuljan et al. (1999),
Famaey et al. (2005), Antoja et al. (2008), Zhao et al. (2009),
Bobylev et al. (2010), etc.

When preparing Table 3, we compared the coordinates of the peaks
with those from Dehnen (1998), Zhao et al. (2009), and Antoja et
al. (2012). The presence of the peak detected by Antoja et al.
(2012) is confirmed. Moreover, we detected a previously unknown at
$(U,V)=(-96,-10)\pm(5.1,1.5)$ km s$^{-1}$ and a separate feature
in the Wolf 630 stream at $(U,V)=(30,-40)\pm(3.0,3.0)$ km
s$^{-1}$.

In general, we can conclude that there is an important difference
of our $UV$ velocity distribution from the previously known ones
constructed from a smaller amount of data from the Hipparcos
catalogue. This difference lies in the presence of an extremely
developed branch in the region of the Wolf~630 peak. It can be
seen from Fig. 3 that the boundary of this region extends up to
$U\approx70$ km s$^{-1}$. When other data are analyzed, this
boundary usually did not extend farther than $U\approx50$ km
s$^{-1}$ (Dehnen 1998; Bobylev et al. 2010; Bubar and King 2010;
Antoja et al. 2012).

It follows from Fig. 3 that the $UW$ velocity ellipsoid is
parallel to the Galactic plane, The $VW$ velocity distribution is
bimodal, because the Hercules stream is very powerful and, hence,
forms a separate clump here. No such effect was observed on the
corresponding map constructed from Hipparcos data (Dehnen 1998;
Bobylev et al. 2010).

\section*{CONCLUSIONS}
We considered stars from the RAVE4 catalogue with known radial
velocities, proper motions, and distance estimates. We showed that
the solar peculiar velocity components $U_\odot, V_\odot,
W_\odot,$ the angular velocity of Galactic rotation at the solar
distance $\Omega_0,$ and its first derivative $\Omega'_0$ are
satisfactorily determined from stars within 3 kpc of the Sun. The
second derivative $\Omega''_0$ is determined well only from
samples of distant stars at distances of 0.5--3 kpc; $\Omega_0$
decreases in this case. The difficulties (the impossibility to
simultaneously determine all the parameters of interest to us) are
related to the complex space distribution of objects from the RAVE
catalogue.

We obtained one of the solutions based on a sample of 183 631
stars with ages in the interval $9.4<\lg t<9.8:$
 $(U,V,W)_\odot=(10.46, 21.48, 7.19)\pm(0.11,0.09,0.08)$ km s$^{-1}$;
   $\Omega_0=24.36\pm0.13$ km s$^{-1}$ kpc$^{-1}$;
  $\Omega'_0=-3.33\pm0.03$ km s$^{-1}$ kpc$^{-2}$;
 $\Omega''_0= 0.11\pm0.05$ km s$^{-1}$ kpc$^{-3}$.
This gives the linear rotation velocity of the Sun’s around the
Galactic center $V_0=195\pm10$ km s$^{-1}$ for the adopted
$R_0=8.0\pm0.4$ kpc.

When using nearer stars having, on average, smaller errors in
their space velocities, the rotation velocity of the solar
neighborhood around the Galactic center turns out to be higher. We
found the following kinematic parameters from a sample of 145 807
stars at distances $r<0.5$~kpc:
 $(U,V,W)_\odot=(9.12,20.80,7.66)\pm(0.10,0.10,0.08)$ km s$^{-1}$,
   $\Omega_0=28.71\pm0.63$ km s$^{-1}$ kpc$^{-1}$, and
 $\Omega'_0= -4.28\pm0.11$ km s$^{-1}$ kpc$^{-2}$. This
gives $V_0=230\pm12$ km s$^{-1}$ (for $R_0=8.0\pm0.4$ kpc),
 $A= 17.12\pm0.45$ km s$^{-1}$ kpc$^{-1}$, and
 $B=-11.60\pm0.77$ km s$^{-1}$ kpc$^{-1}$.

It was noticed that the error per unit weight $\sigma_0$ increased
considerably at an age $\lg t<8.5,$ reaching approximately 35 km
s$^{-1}$. This is larger than the expected error for stars of such
an age, about 25 km s$^{-1}$. This means that either the distance
estimates or the ages are not reliable for the youngest stars from
the RAVE catalogue.

We constructed the 2D velocity distributions in the $UV,$ $UW,$
and $VW$ planes. The $UV$ velocity distribution differs from the
previously known distributions constructed from a smaller amount
of data from the Hipparcos catalogue. This difference lies in the
fact that our distribution shows an extremely enhanced branch in
the Wolf 630 region. We detected a previously unknown stream at
 $(U,V)=(-96,-10)$ km s$^{-1}$ and a separate new feature in the
Wolf~630 stream with coordinates $(U,V)=(30,-40)$ km s$^{-1}$.

\subsection*{ACKNOWLEDGMENTS}
We are grateful to the referees for their useful remarks that
contributed to an improvement of our paper.

\subsection*{REFERENCES}
\quad~{\small
1. T. Antoja, F. Figueras, D. Fern\'andez, and J.
Torra, Astron. Astrophys. 490, 135 (2008).

2. T. Antoja, A. Helmi, O. Bienaym\'e, J. Bland-Hawthorn, B.
Famaey, K. Freeman, B.K. Gibson, G. Gilmore, E.K. Grebel, et al.,
Mon. Not. R. Astron. Soc. 426, L1 (2012).

3. J. Binney, B. Burnett, G. Kordopatis, P.J. McMillan, S. Sharma,
T. Zwitter, O. Bienaym\'e, J. Bland-Hawthorn, et al., Mon. Not. R.
Astron. Soc. 437, 351 (2014a).

4. J. Binney, B. Burnett, G. Kordopatis, M. Steinmetz, G. Gilmore,
O. Bienaym\'e, J. Bland-Hawthorn, B. Famaey, et al., Mon. Not. R.
Astron. Soc. 439, 1231 (2014b).

5. V.V. Bobylev and A.T. Bajkova, Astron. Rep. 51, 372 (2007).

6. V.V. Bobylev, A.T. Bajkova, and A.A. Myll\"ari, Astron. Lett.
36, 27 (2010).

7. V.V. Bobylev and A.T. Bajkova, Mon. Not. R. Astron. Soc. 441,
142 (2014).

8. M.A. Breddels, M.C. Smith, A. Helmi, O. Bienaym\'e, J. Binney,
J. Bland-Hawthorn, C.~Boeche, B.C.M. Burnett, et al., Astron.
Astrophys. 511, A90 (2010).

9. E.J. Bubar and J.R. King, Astron. J. 140, 293 (2010).

10. B. Burnett and J. Binney, Mon. Not. R. Astron. Soc. 407, 339
(2010).

11. B. Burnett, J. Binney, S. Sharma, M. Williams, T. Zwitter, O.
Bienaym\'e, J. Bland-Hawthorn, K.C. Freeman, et al., Astron.
Astrophys. 532, A113 (2011).

12. D. Chakrabarty, Astron. Astrophys. 467, 145 (2007).

13. C.K. Chui, Wavelets: A Mathematical Tool for Signal Analysis
(SIAM, Philadelphia, PA, 1997).

14. B. Co\c{c}kuno\u{g}lu, S. Ak, S. Bilir, S. Karaali, E. Yaz, G.
Gilmore, G.M. Seabroke, O.~Bienaym\'e, et al., Mon. Not. R.
Astron. Soc. 412, A1237 (2011).

15. W. Dehnen, Astron. J. 115, 2384 (1998).

16. W. Dehnen, Astrophys. J. 524, L35 (1999).

17. W. Dehnen, Astron. J. 119, 800 (2000).

18. B. Famaey, A. Jorissen, X. Luri, M. Mayor, S. Udry, H.
Dejonghe, and C. Turon, Astron. Astrophys. 430, 165 (2005).

19. T. Foster, and B. Cooper, The Dynamic Interstellar Medium: A
Celebration of the Canadian Galactic Plane Survey, eds R. Kothes,
T.L. Landecker, A.G. Willis, ASP Conf. Ser. 438, 16 (2010).

20. C. Francis and E. Anderson, New Astron. 14, 615 (2009).

21. R. Fux, Astron. Astrophys. 373, 511 (2001).

22. E. Gardner and C. Flinn, Mon. Not. R. Astron. Soc. 405, 545
(2010).

23. G.A. Gontcharov, Astron. Lett. 32, 759 (2006).

24. The HIPPARCOS and Tycho Catalogues, ESA SP--1200 (1997).

25. J. Holmberg, B. Nordstr\"om, and J. Andersen, Astron.
Astrophys. 475, 519 (2007).

26. J. Holmberg, B. Nordstr\"om, and J. Andersen, Astron.
Astrophys. 501, 941 (2009).

27. M. Honma, T. Nagayama, K. Ando, T. Bushimata, Y.K. Choi, T.
Handa, T. Hirota, H.~Imai, T. Jike, et al., Publ. Astron. Soc.
Jpn. 64, 136 (2012).

28. S. Karaali, S. Bilir, S. Ak, E. Yaz G\"ok\c{c}e, \"O. \"Onal,
and T. Ak, Publ. Astron. Soc. Austral. 31, 13 (2014).

29. R. Klement, B. Fuchs, and H.-W. Rix, Astrophys. J. 685, 261
(2008).

30. G. Kordopatis, G. Gilmore, M. Steinmetz, C. Boeche, G.M.
Seabroke, A. Siebert, T.~Zwitter, J. Binney, P. de Laverny, et
al., Astron. J. 146, A134 (2013).

31. F. van Leeuwen, Astron. Astrophys. 474, 653 (2007).

32. S. Pasetto, E.K. Grebel, T. Zwitter, C. Chiosi, G. Bertelli,
O. Bienaym\'e, G. Seabroke, J. Bland-Hawthorn, et al., Astron.
Astrophys. 547, A71 (2012).

33. M.J. Reid, K.M. Menten, A. Brunthaler, X.W. Zheng, T.M. Dame,
Y. Xu, Y. Wu, B.~Zhang, et al., Astrophys. J. 783, 130 (2014).

34. R. Sch\"onrich, J. Binney, and W. Dehnen, Mon. Not. R. Astron.
Soc. 403, 1829 (2010).

35. A. Siebert, M.E.K. Williams, A. Siviero, W. Reid, C. Boeche,
M. Steinmetz, J. Fulbright, U. Munari, T. Zwitter, et al., Astron.
J. 141, 187 (2011).

36. J. Skuljan, J.B. Hearnshaw, and P.L. Cottrell, Mon. Not. R.
Astron. Soc. 308, 731 (1999).

37. M. Steinmetz, T. Zwitter, A. Seibert, F.G. Watson, K.C.
Freeman, U. Munari, R.~Campbell, M. Williams, et al., Astron. J.
132, 1645 (2006).

38. V.V. Vityazev, Wavelet-Analysis of Time Series (SPb. Gos.
Univ., St.-Petersburg, 2001) [in Russian].

39. M.E.K. Williams, M. Steinmetz, J. Binney, A. Siebert, H. Enke,
B. Famaey, I. Minchev, R.S. de Jong, et al., Mon. Not. R. Astron.
Soc. 436, 101 (2013).

40. N. Zacharias, C. Finch, T. Girard, A. Henden, J.L. Bartlett,
D.G. Monet, and M.I. Zacharias, Astron. J. 145, 44 (2013).

41. J. Zhao, G. Zhao, and Y. Chen, Astrophys. J. 692, L113 (2009).

42. T. Zwitter, A. Siebert, U. Munari, K.C. Freeman, A. Siviero,
F.G. Watson, J.P. Fulbright, R.F.G. Wyse, et al., Astron. J. 136,
421 (2008).

43. T. Zwitter, G. Matijevi\v{c}, M.A. Breddels, M.C. Smith, A.
Helmi, U. Munari, O. Bienaym\'e, J. Binney, J. Bland-Hawthorn, et
al., Astron. Astrophys. 522, A54 (2010). }

\end{document}